\documentclass{article}
\usepackage[utf8]{inputenc}
\usepackage[a4paper, left=1cm, right=1cm, top=1cm, bottom=2cm]{geometry}
\usepackage{graphicx,url,amsmath,amssymb,titlesec,hyperref} % hyperref must be after titlesec
\usepackage[version=3]{mhchem}
\hypersetup{hidelinks}
\usepackage{textgreek}
\usepackage{authblk}

\title{Neutron imaging of high-temperature Na-Zn Cells: implications for cell design and fabrication}
\author[1]{W. Nash}
\author[1]{M. Sarma}
\author[1]{T. Lappan}
\author[2]{P. Trtik}
\author[3]{C. K. W. Solem}
\author[3]{Z. Wang}
\author[4]{A. Beltrán}
\author[1]{N. Weber}
\author[1]{T. Weier}

\affil[1]{Institute of Fluid Dynamics, Helmholtz-Zentrum Dresden–Rossendorf, Bautzner Landstrasse 400, 01328 Dresden, Germany}
\affil[2]{Laboratory for Neutron Scattering and Imaging, Paul Scherrer Institut, Forschungsstrasse 111, 5232 Villigen PSI, Switzerland}
\affil[3]{SINTEF Industry, 7491 Trondheim, Norway}
\affil[4]{Instituto de Investigaciones en Materiales, Unidad Morelia, Universidad Nacional Autónoma de México (UNAM), Antigua Carretera a Pátzcuaro No. 8701, Col. Ex Hacienda de San José de la
Huerta, C.P. 58190 Morelia, Michoacán, Mexico}

\date{}

\begin{document}

\maketitle
\tableofcontents

\section*{Abstract}
Electrochemical cells employing Sodium (Na) and Zinc (Zn) electrodes and a chloride salt electrolyte have been imaged by neutron radiography during cycling.  The use of such abundant raw materials confers a very low energy-normalised cost to the Na-Zn system, but its operation requires them to be entirely molten, and therefore to be operated at 600\textdegree C.  To suppress the self-discharge that results from this all-molten configuration, porous ceramic diaphragms are used to partition the electrolyte and thereby impede the movement of the Zn$^{2+}$ ions responsible towards the Na electrode.  Neutron images reveal large gas bubbles trapped beneath these diaphragms, formed during the cell fabrication process due to the large volume change that accompanies melting/solidifying of the electrolyte.  Cycling data confirm that these bubbles interfere with cell operation by substantially increasing ohmic resistance.  They indicate the need for either a new diaphragm design, or a cell fabrication process that prevents their formation in the first instance.

\section{Introduction and Motivation}
Replacing fossil fuel-derived energy with intermittent renewables will require reductions in the cost of batteries \cite{iea2024}.  Fully molten Na-Zn cells \cite{xuNaZnLiquidMetal2016} theoretically offer unprecedented reductions in cost, due to their comparatively high discharge voltage (1.8V) and abundant constituent materials.  While the 600\textdegree C operating temperature does pose significant engineering challenges for the Na-Zn system, it does not significantly reduce its efficiency for the envisioned use-case, which is large-scale grid-backup rather than behind-the-meter storage for individual homesteads.  Research interest in the Na-Zn system is further motivated by its low environmental footprint, non-toxic components, and immunity to thermal runaway \cite{weberetal2024}.  Nonetheless, the technology presently remains at an early stage of development.  With one exception \cite{sarmaReusableCellDesign2024a}, all Na-Zn cells constructed to date have been unenclosed, and operated inside Ar-filled gloveboxes to prevent oxidation of their Na electrodes \cite{xuNaZnLiquidMetal2016} \cite{xuElectrodeBehaviorsNaZn2017} \cite{zhang2022} \cite{godinez-brizuela2023}.  The development of enclosed cells, equipped with gas-tight seals that enable them to be operated outside a laboratory, is a necessary prerequisite for commercialization.  The chemical processes occurring in such cells are difficult to study however, due to the opacity of their enclosing capsules/lids.  In particular, flow structures in their electrolytes, or growth/deformation of their electrodes, cannot be observed.  Neutron-beam radiography overcomes this problem, thereby enabling the performance characteristics of closed cells to be better understood.

\section{Cell Design and Operating Principles}
The Na-Zn cell comprises three completely molten phases (see Figure 1a): a Na negative electrode (top); a NaCl-CaCl$_2$-ZnCl$_2$ salt electrolyte (middle); and a Zn positive electrode (bottom).  This layered structure is maintained by these phases' progressively higher densities and mutual immiscibility.  The discharge process comprises the coupled oxidation of Na and reduction of Zn$^{2+}$; the latter species being displaced by Na$^{+}$ ions in the electrolyte.  Combining these redox processes yields the overall cell reaction: 

\begin{align}
\cee{2Na + ZnCl_2 &<=> 2NaCl + Zn.}
\end{align}

The cell's relatively high 1.8V open-circuit voltage arises from the large difference between the decomposition potentials of NaCl and ZnCl$_2$ \cite{hamer1956}.  Successful exploitation of this reaction however depends on minimizing contact between the Na electrode and the Zn$^{2+}$ ions in the electrolyte, since direct exchange of electrons between these species -- rather than via the cell's external circuit -- constitutes self-discharge and wasted energy.  An essential structural component of a Na-Zn cell is therefore the porous ceramic diaphragm that partitions the electrolyte into upper ('anolyte') and lower ('catholyte') layers.  By permitting only diffusive mass transport between these layers, and preventing comparatively rapid processes of advection, this diaphragm is intended to ensure that Zn$^{2+}$ ions formed during charge are mostly retained in the catholye \cite{duczeketal2024}.  However, the efficacy of this diaphragm, and its influence on other aspects of cell operation, remains one of the least experimentally-studied aspects of the Na-Zn cells to date, and is therefore a major focus of the present study.

\begin{figure}[h!]
    \centering
    \includegraphics[width=0.9\textwidth]{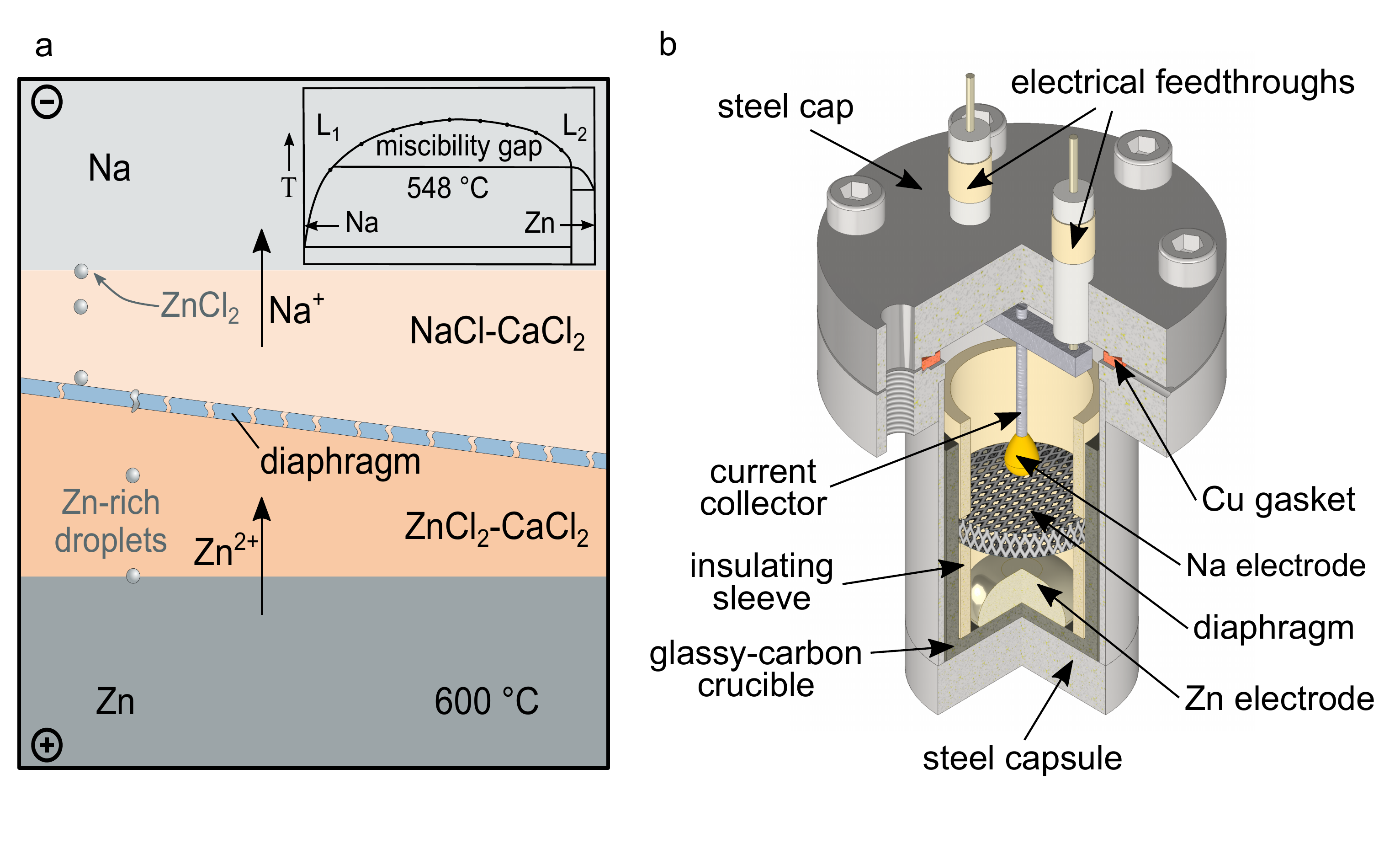}
    \caption{a) Electrochemical operating principles of a Na-Zn cell.  Note the difference in chemical composition between the electrolyte's upper and lower layers (orange) maintained by the porous ceramic diaphragm (pale blue).  Note also the Zn droplets formed by contact between ZnCl$_2$ and the Na electrode (a biproduct of self-discharge).  The Na-Zn binary phase diagram \cite{cetin1991} -- showing the extent of Na-Zn immiscibility -- is shown in as inset (top-right corner).  b) Engineering diagram of the Na-Zn cell prototype used for neutron imaging in this study.  The electrolyte (not shown) fills the space between the current collector and the Zn electrode. Current collectors were either needle-shaped (as shown here) or disc-shaped, sometimes with ceramic 'Na collectors' capping them.}
\end{figure}

The aforedescribed self-discharge mechanism imposes an additional performance requirement on the diaphragm. Since \textit{total} retention of Zn$^{2+}$ ions in the catholyte is not achievable, some metallic Zn will inevitably form at the Na electrode and must be returned to the Zn electrode.  Due to the immiscibility between Na and Zn which prevails at the cell's 600\textdegree C operating temperature \cite{cetin1991}, Zn should form droplets that detach from the Na electrode and sink through the electrolyte to the Zn electrode, driven by their high density (see Figure 1a).  This however, requires the diaphragm to have pores that are large enough to enable the passage of molten Zn droplets.  The diaphragm's pore size is therefore a critical parameter, which must achieve a compromise between the suppression of self-discharge and the possibility of Zn recovery.  To search for this compromise, cells containing diaphragms with different pore sizes were investigated in this study.

\section{Methodology}
Neutron radiography is an appropriate technique for imaging Na-Zn cells because of the large differences in neutron attenuation coefficient exhibited by their constituent elements.  For thermal neutrons with a mean energy of 25 meV, these coefficients are 0.09 cm$^{-1}$ for Na, 0.08 cm$^{-1}$ for Ca, 0.35 cm$^{-1}$ for Zn, and 1.33 cm$^{-1}$ for Cl \cite{NEA1994}.  The metallic electrodes and metal-chloride electrolyte can therefore be readily distinguished.  Zinc droplets formed by self-discharge can also be resolved, provided these are larger than the achievable (on average) 0.1 mm spatial resolution.

Acquisition of good-contrast images necessitates the use of very thin or low neutron-absorbing cell housing materials.  Custom-designed cells and furnaces were developed specifically for this purpose, the latter employing ceramic heating plates (positioned beneath the cell so as not to obscure the neutron beam) and porous calcium silicate insulation.  The cell design employed is shown schematically in Figure 1b.  Note that the glassy carbon crucible containing the electroactive materials, together with the surrounding steel capsule, served as the positive current collector.  A more detailed description of the cell, specifying aspects such as its capsule materials and feedthrough design, is provided by \cite{sarmaReusableCellDesign2024a}.  Heating plates were purchased from BACH Resister Ceramics. 

Cells of two different sizes were constructed, so that the influence of electrolyte flow phenomena (typically scale-dependent) on the efficacy of the diaphragm could be investigated.  The cylindrical volumes of electro-active materials in the small and large cells had diameters of 23 mm and 30 mm respectively.  Alumina ceramic foams were used as diaphragms, similar to those used by the aluminum refining industry \cite{kennedy2013}.  These are generally classified by their pore density, commonly quoted in units of pores per inch (PPI).  40 PPI diaphragms were used for the small cells, while 30 PPI and 80 PPI diaphragms were used in the larger cells.  These exhibited average pore diameters of 1.5 mm, 2.5 mm and 0.5 mm, respectively.  The diaphragms were orientated horizontally in the cells as shown in Figure 1b, not sloping as depicted in Figure 1a.

All cells were assembled in an Ar-filled glovebox in the discharged state; i.e. with no metallic Na layer, and with an electrolyte formed from a near-equimolar mixture of NaCl and CaCl$_2$ (NaCl:CaCl$_2$ mole fractions ranged from 0.56:0.44 to 0.44:0.56).  Cell assembly was performed at Helmholtz-Zentrum Dresden Rossendorf in Germany.  Alumina diaphragms were provided by Lanik, Czech Republic.  Alumina wool was sourced from Polytec PT GmbH, Germany.  Zn powder ($>$99.7 \% purity) was obtained from Acros Organics and premelted under Ar to form solid lumps.  NaCl powder ($>$99.99 \% purity) was purchased from Thermo Scientific.  CaCl$_2$ powder ($>$98 \% purity) was purchased from Supelco.  The salts were dried by heating to 250\textdegree C for $>$48 hours under vacuum.

Cells were cycled using a Biologic SP-150 potentiostat operated in grounded mode.  Temperature was monitored using K-type thermocouples, inserted into shallow holes recessed into the cap of each experimental cell.  Neutron radiography was performed at the NEUTRA facility \cite{lehmann2001} of the Swiss Spallation Neutron Source (SINQ) at the Paul Scherrer Institute in Switzerland. The neutron source was operated with around 1.3 mA proton beam current. The cell was placed in the measuring position 2. $^6$LIF:ZnS scintillator with a 0.2 mm thickness was employed for the radiography studies. The results presented in this manuscript were obtained with a frame rate of 1 frame/sec, and a $100\times100$mm$^2$ field of view.

\section{Discussion}

Figures 2 shows neutron images of a small cell (2a, 40 PPI diaphragm) and a large cell (2b, 30 PPI diaphragm) being charged at 600\textdegree C.  The brightness of each pixel corresponds to the integrated neutron attenuation along the path it represents through the cell (neglecting scattering effects).  Darker pixels indicate greater attenuation.  The darkest regions represent the electrolyte, against which the brighter Zn electrode and diaphragm are clearly visible.  Also visible however, are large gas bubbles located beneath the diaphragms.  These are present in every cell, and in most cells -- including all of the small ones -- they are large enough to entirely cover the lower surface of the diaphragm.  Attempts to encourage their passage through the diaphragm by lateral percussion with a hammer were unsuccessful.

Cycling data demonstrate that these bubbles detrimentally affect cell performance.  Significantly greater potentials are required to charge cells with bubbles covering their entire diaphragm than for cells with only partially obstructing bubbles at the same current density (see Figure 3c).  Charging or discharging at such high potentials damages the cells, by causing electrolytic decomposition of their electrolytes to form chlorine gas.  This can be observed in time-lapse videos of neutron images of the smaller cells.  Figure 2d shows two screenshots from one such video (provided in full in the supplementary materials).  Transient bright regions at the perimeter of the diaphragm are accompanied by rapid changes in the surface level of the salt.  This is strongly indicative of gas forming at the salt's interface with the glassy-carbon crucible (now acting as a current collector), disturbing the upper surface of the salt as it rises and escapes.  Commercial operation of Na-Zn cells at such excessive voltages would be prohibited by safety limits, which would limit bubble-affected cells to very low current densities.  

The utility of neutron imaging for gas bubble detection is potentially of interest to development efforts for conventional all-liquid batteries, since these frequently employ foam current collectors -- very similar in structure to the diaphragms investigated here -- submerged in molten mixtures of halides \cite{kelley&weier2018}.    

\begin{figure}[h!]
    \centering
    \includegraphics[width=0.8\textwidth]{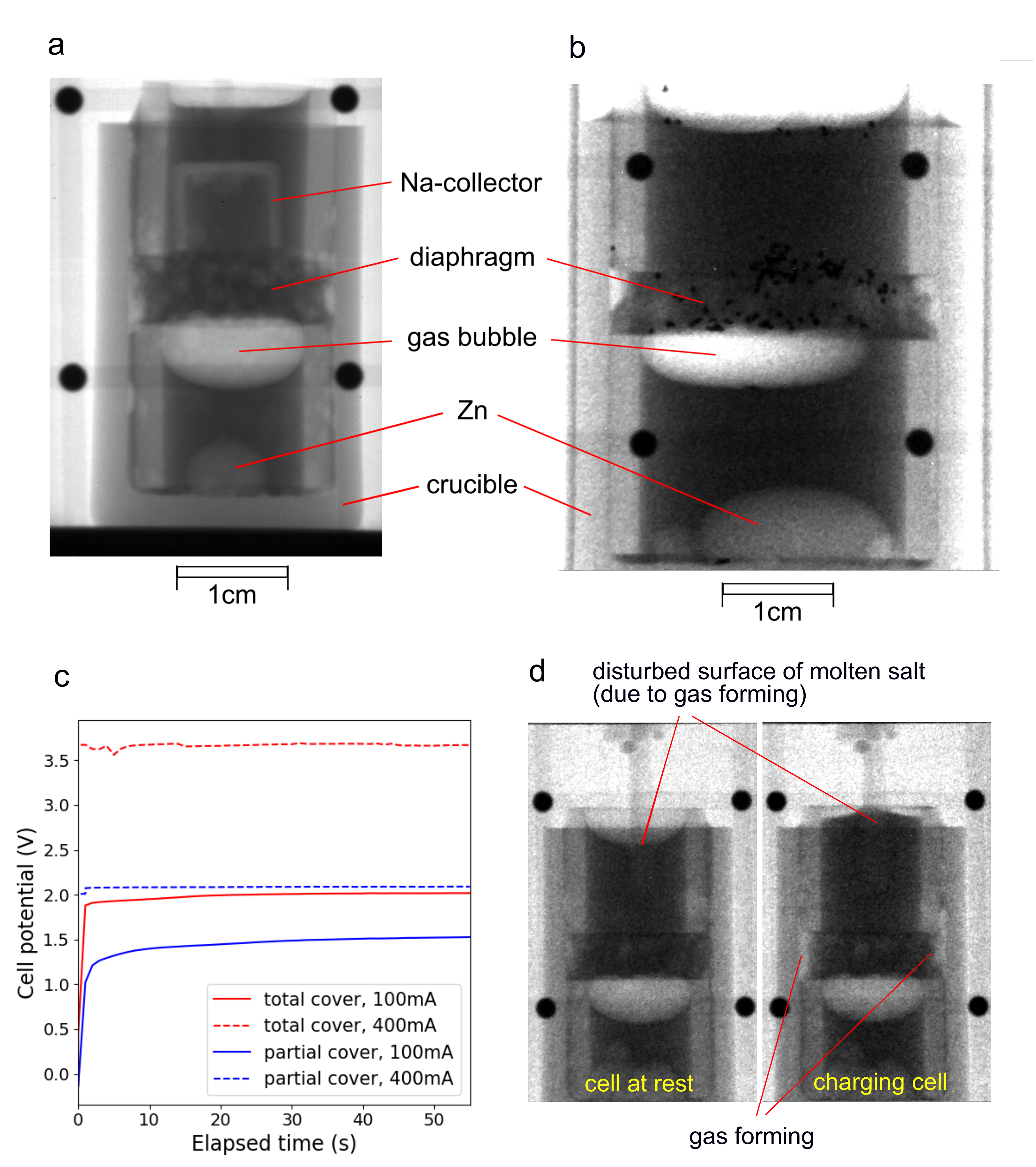}
    \caption{a) Neutron image of small Na-Zn cell during cycling.  b) Neutron image of large Na-Zn cell during charging.  The very dark circles connected by a pale-grey mesh are part of the gadolinium black body grid \cite{Boillat201815769} that are usually used to quantify the effect of scattered neutrons.  c) Comparison between the potentials needed to charge cells with bubbles that totally (red) and partially (blue) cover their diaphragms, at 100mA and 400mA.  d) Neutron images of a small Na-Zn cell when resting (left) and charging (right).  The evolution of gas from the electrolyte where it makes contact with the glassy carbon current collector is evident.}
\end{figure}

\subsection{Origin of Bubbles}
Timelapses of neutron images obtained during the heating of the cells indicate that the bubbles do not originate from degassing of the salt.  Rather, they are present at the outset, and must therefore form during cell fabrication.  Fabrication entails the pouring of molten salt into the crucible, after which the diaphragm and sleeves (see Figure 1b) are forcefully submerged beneath it using an alumina rod.  The mechanical agitation involved in submerging the diaphragm removes bubbles that are trapped beneath it, and this can be confirmed by visual inspection, facilitated by the transparency of the molten salt.  The gas bubbles observed in the neutron images must therefore form during the solidification of the salt, most likely as a result of the decrease in volume that accompanies its crystallization.

Experience handling molten salts in the laboratory demonstrates the difficulty inherent in controlling their shape upon cooling.  By the standards of most other common substances, chloride salts undergo extreme volume changes upon solidification.  For the near-equimolar NaCl-CaCl$_2$ mixtures used in Na-Zn cells, the decrease in volume is greater than 9 \% (the densities of the melt and solid being 1.97 g/cm$^{3}$ \cite{janz1988}\cite{weier2024} and 2.16 g/cm$^{3}$ \cite{crc2015} respectively).  Bodies of crystallized salt generally contain large cavities, formed because solidification first occurs at the cooling volume's surface before proceeding inward.  The volume of the bubbles -- estimated from the neutron images by modelling them as simple surfaces of revolution -- is consistent with a 9 \% contraction of the total salt volume present.  The formation of the bubbles is therefore readily explained.  Their inability to rise through the diaphragm upon remelting of the salt reflects its high surface-tension, and the negligible difference in pressure between the gas inside the bubbles and the gas overlying the electrolyte.

\subsection{Options for bubble mitigation}
Elimination of gas bubbles can be attempted by either modifying the cell design to facilitate their escape, or by changing the fabrication procedure to prevent their formation in the first instance.  Of the strategies belonging to the latter category, perhaps the most elegant is to cool the newly-fabricated cells from below so the salt freezes from the base upwards.  Cavities formed by contracting salt can then be filled by overlying salt that is still molten.  The fact that the larger cells imaged sometimes contain smaller bubbles (in relative rather than absolute terms) lends support to this approach, since it demonstrates that bubble size is sensitive to cooling rate.  It is possible that simply cooling more slowly -- as occurred in the larger cells due to their greater size -- can eliminate the bubbles.  These approaches certainly deserve attention, but their success cannot be guaranteed at the outset because of the complex pore structure of the diaphragm, which might locally interfere with the rate of crystallization and the accessibility of cavities to overlying molten salt.  Furthermore, such interference is only likely to intensify for the smaller pore sizes that are better suited to controlling Zn$^{2+}$ migration (see section 5.3).

An alternative modification of the fabrication procedure -- involving more complex equipment -- is to assemble the cells under vacuum, or to withdraw Ar from inside them using a vacuum pump prior to cycling, mechanically drawing any bubbles present through the diaphragm and out of the system.  Avoidance of such complex procedures is desirable however, if the Na-Zn system is to offer production costs that are commercially competitive.

If modifications of the fabrication procedure prove unsuccessful, the cell must be redesigned so bubbles are able to escape from beneath the diaphragm.  Such changes are probably necessary in any case, since it is desirable that future Na-Zn storage facilities be tolerant to unexpected disruptions to their operation that cool their cells and freeze their electrolytes.  One possible design modification is a change in the chemical composition of the electrolyte.  NaCl must of course remain as a solution component, since it supplies essential Na$^{+}$ to the negative electrode.  But CaCl$_2$ could be replaced with a halide that forms -- together with NaCl -- a solution that undergoes a smaller volume change upon crystallization.  The choice of alternatives to CaCl$_2$ is restricted however by the requirement that the new solution have a acceptably low melting temperature, and that the new component not participate in the cell's redox reactions.  SrCl$_2$ is suitably inert and low-melting, but negligibly reduces the volume change upon crystallization \cite{janz1988}.  BaCl$_2$ reduces it to 4 \%\cite{janz1988} and is also suitably stable, but its toxicity is unwelcome.  The decomposition potentials for LiCl and KCl indicate that they are probably electrochemically compatible, but the solutions they form with NaCl contract even more than those containing CaCl$_2$ (36\% and 29\% respectively \cite{crc2015} \cite{janz1988}).  If no suitable alternative salt mixtures can be found, then the cell component that must be redesigned is the diaphragm. 

\subsection{Possible implications for diaphragm design}

Enlarged pores or dedicated holes might enable bubbles to pass through the diaphragm, however these modifications must not undermine its capacity to suppress the upward transport of Zn$^{2+}$ ions.  Both options involve changing the porosity; defined for the purpose of this calculation as the ratio of the pore area in a given horizontal section through the diaphragm to its overall cross-sectional area.  To help explore the range of permissible design options, the sensitivity of self-discharge rate to the diaphragm's porosity and thickness are estimated in Figure 3 for the larger of the cells imaged in this study (approx. 1Ah total capacity).  The black and grey lines represent diaphragms with thicknesses of 0.5 cm and 5 cm respectively, for a range of porosities (individually labelled).  The blue line represents the diaphragms employed in the cells imaged in this study (0.5 cm thickness, porosity approx. 0.9).  To assist interpretation, the calculated self-discharge rates are also expressed as the fraction of the cell's total capacity lost per hour, on the figure's right-hand axis.  

The self-discharge rates in Figure 3 are calculated by first estimating Zn$^{2+}$ fluxes through the diaphragm ($J$) using Fick's law:

\begin{equation}
    J = -D \frac{d\varphi_{Zn^{2+}}}{dx},
\end{equation}

for which $D$ denotes the Zn$^{2+}$ diffusion coefficient, $\varphi_{Zn^{2+}}$ denotes the Zn$^{2+}$ concentration, and $x$ denotes the diaphragm thickness.  A uniform concentration gradient is assumed through the diaphragm, with the Zn$^{2+}$ concentrations immediately above and below it fixed at zero and 0.001 mol/cm$^{3}$ respectively; the latter value corresponding to the catholyte composition at full charge. Self-discharge rates ($i$) are then calculated from the Zn$^{2+}$ fluxes by assuming instantaneous mixing within the anolyte and complete and instantaneous reduction of all transferred Zn$^{2+}$ at the Na electrode, by applying Faraday's law:

\begin{equation}
    i = JFv .
\end{equation}

$F$ denotes here the Faraday's constant, and $v$ denotes the valence of Zn$^{2+}$.  Porespace tortuosity, which increases the diaphragm's effective thickness and thereby reduces the Zn$^{2+}$ diffusion rate, is neglected in these calculations because values significantly greater than 1 (negligible tortuosity) are almost certain to exacerbate the problem of bubble retention.  Tortuosity measurements of ceramic diaphragms similar to those used here have yielded values in the range 1.3 to 3.2 \cite{kennedy2013}.

It should be emphasized that the 'worst-case' assumptions made here (instantaneous mixing and Zn$^{2+}$ reduction, negligible tortuosity, fixed Zn$^{2+}$ concentration gradient) yield a conservative measure of self-discharge rate, that is applicable for the specific case of a fully charged cell.  In practice, as the cell discharges and the ZnCl$_2$ concentration in the catholyte decreases, so does the driving force for diffusion, and hence the rate of self-discharge. 

The Zn$^{2+}$ diffusion coefficient (D$_{Zn^{2+}}$) is used as the independent variable in Figure 3 because it is the least constrained parameter.  Measurements of self-diffusion in pure ZnCl$_2$ \cite{sjoblom1968a} have revealed an exponential dependence on temperature (shown in the inset plot of Figure 3), which yields a value of 1.4x10$^{-7}$ cm$^{2}$/s at the Na-Zn cell's 600\textdegree C operating temperature.  This temperature however, lies outside the experimentally investigated interval (329-530\textdegree C; solid portion of the inset plot), making Zn$^{2+}$s diffusion coefficient highly uncertain.  In addition, chemical interactions between Zn$^{2+}$ and the other components in the electrolyte are likely to accelerate or retard its diffusion by creating gradients in chemical potential (so called 'migration').  To account for the resulting uncertainty in self-discharge rate, a three order-of-magnitude range in D$_{Zn^{2+}}$ is represented in Figure 3. 

\begin{figure}[h!]
    \centering
    \includegraphics[width=0.8\textwidth]{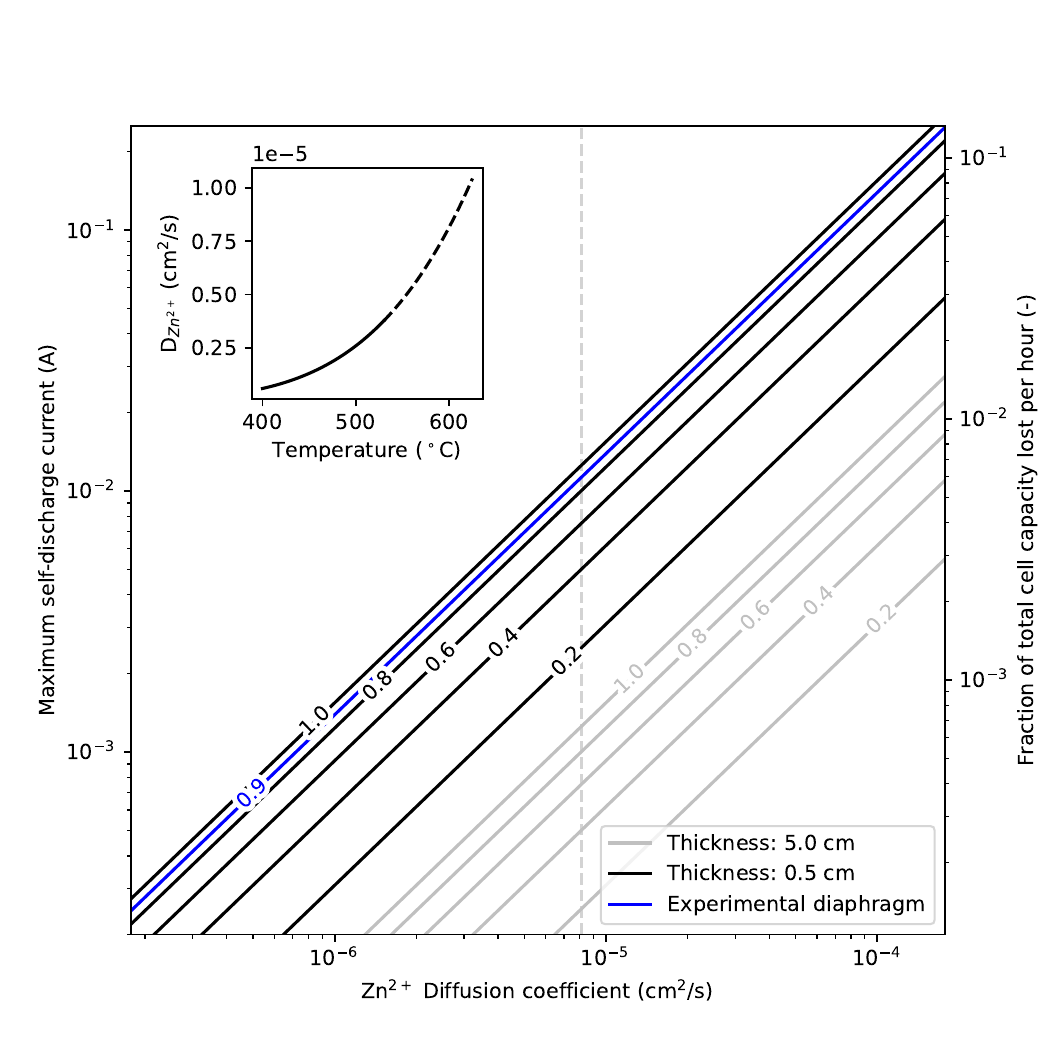}
    \caption{Modelled self-discharge rate for a fully charged cell as a function of diaphragm porosity and thickness.  The vertical dashed line indicates the ZnCl$_2$ self-diffusion coefficient (D$_{Zn^{2+}}$) at 600\textdegree C reported by \cite{sjoblom1968a}.  The temperature dependence of this coefficient, reported by \cite{sjoblom1968a}, is shown in the inset plot.  The dashed section of the curve indicates extrapolation beyond the investigated temperature range.  The blue line (main plot) corresponds to the diaphragm porosity and thickness used in this study.}
\end{figure}

It is clear from Figure 3 that there is only limited scope for increasing diaphragm porosity, since the diaphragms presently employed can be expected to exhibit very rapid self-discharge, as is indeed confirmed by cycling data \cite{sarmaReusableCellDesign2024a}.  On the contrary, porosity \textit{reductions} are needed, in addition to an increase in diaphragm thickness.  This will increase ohmic resistance, however the Na-Zn system's high open-circuit voltage makes this somewhat tolerable.  The experiments presented herein demonstrate that even pores as large as 2.5 mm in diameter will trap bubbles, so further increases in pore-size -- even if compensated for by a reduction in pore density -- are likely to cause unacceptably rapid self-discharge.  

The replacement of pores by a single narrow conduit is possibly a more suitable design, provided it does not induce bulk flow between the anolyte and catholyte.  Such a conduit might require the shape of the diaphragm to deviate from the presently employed disc-shaped geometry, so as to channel bubbles upwards (or Zn droplets downwards) towards it.  Alternatively, the orientation of the diaphragm could be altered in the manner shown in Figure 1a, with two holes positioned off-centre so that bubbles are channeled towards one and Zn droplets towards the other.  The effective porosity of such 'single-hole' diaphragms could be significantly lower than those with many distributed pores, and the corresponding reductions in self-discharge rate are evident from Figure 3.  Alternatively, rather than using a rigid diaphragm a 'paste electrolyte' could be employed, which might enable the upward movement of bubbles while adequately suppressing Zn$^{2+}$ transport.  Paste electrolytes, comprising a mixture of alumina or magnesia powder and molten salt, have been demonstrated in all-liquid bimetallic cells \cite{Cairns1967} \cite{Shimotake1967} \cite{Shimotake1970}.  Pastes employing micron-sized particles, occupying 30-65\% of the volume, exhibit resistances approximately three times greater than that of the pure electrolyte \cite{swinkels1971}.  The high density of Zn compared with magnesia and alumina makes such pastes potentially compatible with the Na-Zn system.

\section{Summary}
This study has demonstrated the utility of neutron radiography for in-situ observation of devices that employ molten metals and salts, such as liquid metal batteries, thermally regenerative cells, reduction cells, and molten-salt reactors/storage devices.  For the specific case of the Na-Zn cells examined here, the presence of gas bubbles beneath their cell diaphragms could not have been confirmed without such an in-situ imaging technique. These bubbles form during the cell fabrication process and subsequently interfere with cell performance.  Their elimination from Na-Zn cell prototypes is a necessary objective that future studies must take measures to achieve. Modification of the fabrication procedure (such as cooling the electrolyte slowly from below) is preferable, however if this is ineffective, the diaphragm must be redesigned or replaced by a paste electrolyte to facilitate bubble escape.  Consideration of these modifications' probable influence on self-discharge emphasizes the need for experimental measurements of the Zn$^{2+}$ diffusion rate in molten NaCl-CaCl$_2$-ZnCl$_2$, to define the limits of acceptable diaphragm/paste porosity, and more broadly, to determine the viability of Na-Zn cells as a viable energy storage technology.

\end{document}